# SECOND LIFE AS A PLATFORM FOR PHYSICS SIMULATIONS AND MICROWORLDS: AN EVALUATION

**Renato P. dos Santos**

ABSTRACT
Often mistakenly seen as a game, the online 3D immersive virtual world Second Life (SL) is itself a huge and sophisticated simulator of an entire Earthlike world. Differently from other metaverses where physical laws are not seriously taken into account, objects created in SL are automatically controlled by a powerful physics engine software. Despite that, it has been used mostly as a mere place for exploration and inquiry, with emphasis on group interaction. This article reports on a study conducted to evaluate the SL environment as a platform for physical simulations and microworlds. It begins by discussing a few relevant features of SL and a few differences found between it and traditional simulators e.g. *Modellus*. Finally, the SL environment as a platform for physical simulations and microworlds is evaluated. Some concrete examples of simulations in SL, including two of our own authorship, will be presented briefly in order to clarify and enrich both discussion and analysis. However, implementation of simulations in SL is not without drawbacks like the lack of experience many teachers have with programming and the differences found between SL Physics and Newtonian Physics. Despite of that, findings suggest it may possible for teachers to overcome these obstacles.

## INTRODUCTION

There are hundreds of *game worlds*, e.g. *World of Warcraft*, specifically created for entertainment, and over 50 different Multi User Virtual Environments (MUVE) currently available (Taylor, 2007), created to simulate real life in some sense. Among them, SL, followed by *OpenSim* and *Active Worlds*, stands out as the platform that offer more services and tools for developing applications with quality (R. Reis, Fonseca, & Escudeiro, 2011) even if it is not the one with the biggest user population (Taylor, 2007). This is the reason why we have chosen it for our research purposes.

## METHODOLOGY

Initially, specific features of the SL environment, relevant to its use as platform for physics microworlds (Papert, 1980) and simulations, will be discussed in detail, as a support for subsequent analysis.

Afterwards, the SL environment as a platform for physics microworlds and simulations will be evaluated, using the criteria established by Reis & Andrade Neto (2002). We will focus on features relevant to the teaching and learning than on others more relevant in terms of computing.
A few concrete examples of simulations in SL, including two of our own authorship, will be presented briefly in order to clarify and enrich both discussion and analysis.

**RELEVANT FEATURES OF THE SL ENVIRONMENT**

On February 2012, the world of SL was made up of thousands of uniquely named *Regions* (Shepherd, 2012) - virtual parcels of land representing an area of 256 m x 256 m - linked together to make the *Maingrid*, *Private*, *Homestead* and *Openspace* areas that spread over 1,998.26 km$^2$ (Shepherd, 2012) of virtual land, not much smaller than the country of Luxembourg. Each Region is rendered by a single process running on Linden Lab servers ("Simulator," 2010) that simulates a full rigid-body physical dynamics, including gravity, elasticity, and conservation of momentum ("Physics engine," 2008), and an accurate, polygon-level collision detection of all the objects in the column-space above its land.

Each server is edge-connected to four other machines as a grid of computers. This grid computes a simplified solution of the Navier-Stokes equations to simulate the motion of winds and clouds that time-evolve across the entire world (Ondrejka, 2004a; Rosedale & Ondrejka, 2003). It also runs user's scripts and streaming routines to send back all the data needed to view the world to anyone's client who is connected (Ondrejka, 2004a). As a result, the SL 'Sun' usually rises and sets each 4 Earth hours always directly opposite a full Moon ("llGetSunDirection," 2009), objects fall under the effect of gravity, trees and grass blow in the wind and clouds form and drift (Ondrejka, 2004a). Therefore, SL is a big simulation of an entire Earthlike world.

Once logged through the client software, SL users can enjoy the 3D scenery, fly, drive cars, interact with other avatars, play games, or create objects. In fact, well over 99% of the objects in SL are user created (Ondrejka, 2004b), what has been characterized as a shift of culture, from a media consumer culture to a participatory culture (Jenkins, Clinton, Purushotma, Robison, & Weigel, 2006).

In a rapid tour through many virtual spaces dedicated to Science, however, one will almost only find real-world replica, mere institutional presences, with traditional classrooms, virtual boards exhibiting 2D PowerPoint™ presentations, and video seminars. We agree with Kapp (2007) that this is certainly not the best use of Second Life or any other metaverse. As we will see below, the real advantage is using it to do innovative things that could not otherwise be done in a classroom that reach into the pupil's imaginations (de Freitas & Griffiths, 2009).

Differently from other virtual environments, SL offers interactivity features to add to the objects through its *Linden Scripting Language* (LSL) ("LSL Portal," n.d.). Its structure is based on *Java* and *C* languages and it provides nearly four hundred functions, among which several of interest to Physics studies in SL used e.g. to access the position or the velocity of the object in the region or to apply a force to it.

However, some important points should be taken into account when one wants to build a simulator or a microworld in SL. See also (dos Santos, 2009) for a deeper discussion on these points.

Time can be affected by different kinds of lag and, therefore, SL will definitely not give response time down to milliseconds consistently for quantitative simulations. There are no fluids in SL (dos Santos, 2009): outside the SL 'oceans', 'water' is a mere texture applicable to an object. Besides, friction, air and water resistance, and buoyancy were not fully implemented in SL. Remarkably enough, light cannot be contained by the walls of a room and, as a consequence, light simply "is there" in SL, as an state of illumination, without any physical mechanism involved in its production or propagation, as in the most primitive conceptions of light (Lindberg, 1981). Finally, SL does not even try to simulate any physics beyond Mechanics, and, therefore, a designer may have to use creativity to simulate other interactions like electromagnetic or nuclear ones.

We have already demonstrated that SL does not simulate a real-world Physics virtualization or the Galilean/Newtonian "idealized" Physics (dos Santos, 2009). It was also noticed that certain physical quantities such as mass, acceleration, and energy have a very different meaning in SL as compared to real-world Physics.

It was further noticed that SL is lacking of some basic functions to provide initial conditions to objects, e.g. their initial velocity and position, as simulators usually have (dos Santos, 2009). We have mentioned there that to circumvent this limitation, in certain circumstances, the designer can use the *llRezObject* function that sets the position and velocity of the object when rezzed. However, physical objects are, in principle, affected by SL 'wind' and 'gravity' and, therefore, they will not keep that velocity constant and will deviate from its rectilinear trajectory. This difficulty can be seen in the following three concrete examples.

*Oddprofessor Snoodle*, a Physics teacher, was recently creating a replica of an *air track* for her *Museum and Science Center* (Figure 1). She found herself unable to have the glider to move along a track at a given constant velocity even if using buoyancy to cut down on friction.

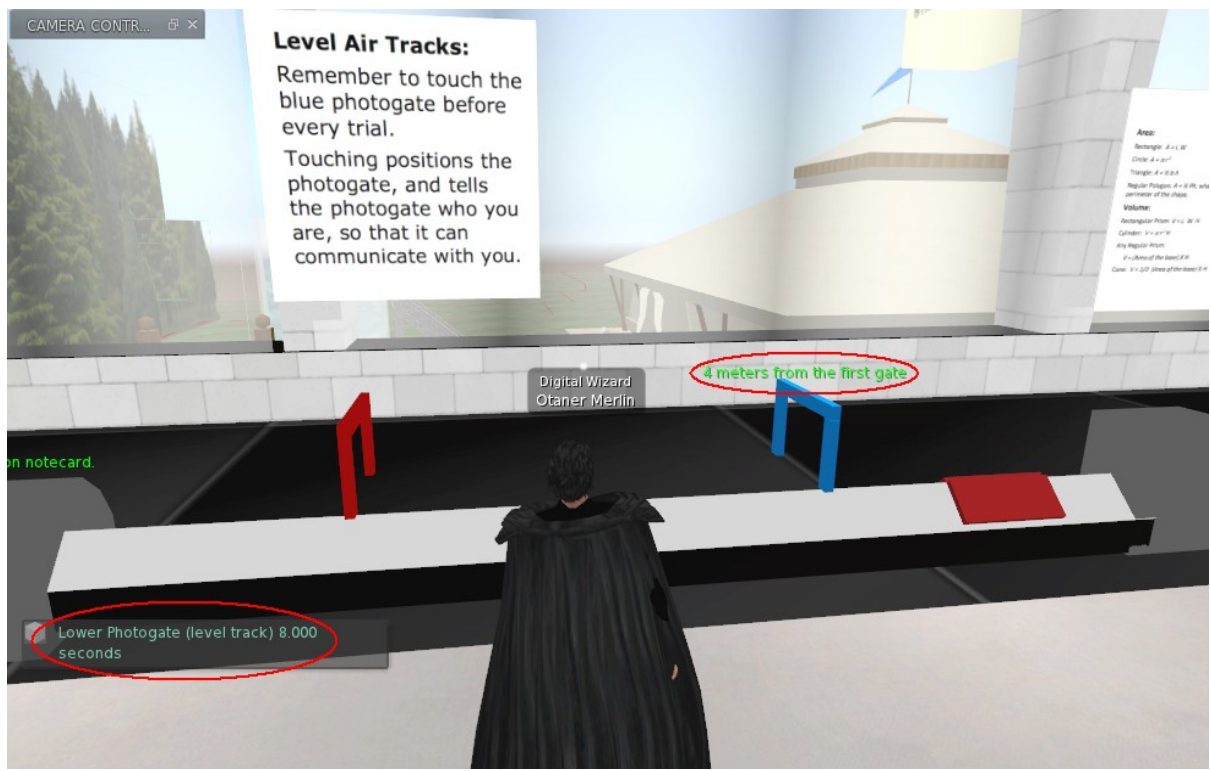

**Figure 1 - Oddprofessor Snoodle's air track replica at the *Museum and Science Center***
Source: photo taken by the author

As another example, Greis & Reategui (2010) have built a simulator themed as a pair of amusement park non-wheeled bumper cars running on rail tracks and performing head-on collisions (Figure 2). In a private communication to this author, Greis manifested the same kind of difficulties in controlling the velocity in SL as *Oddprofessor Snoodle* faced.

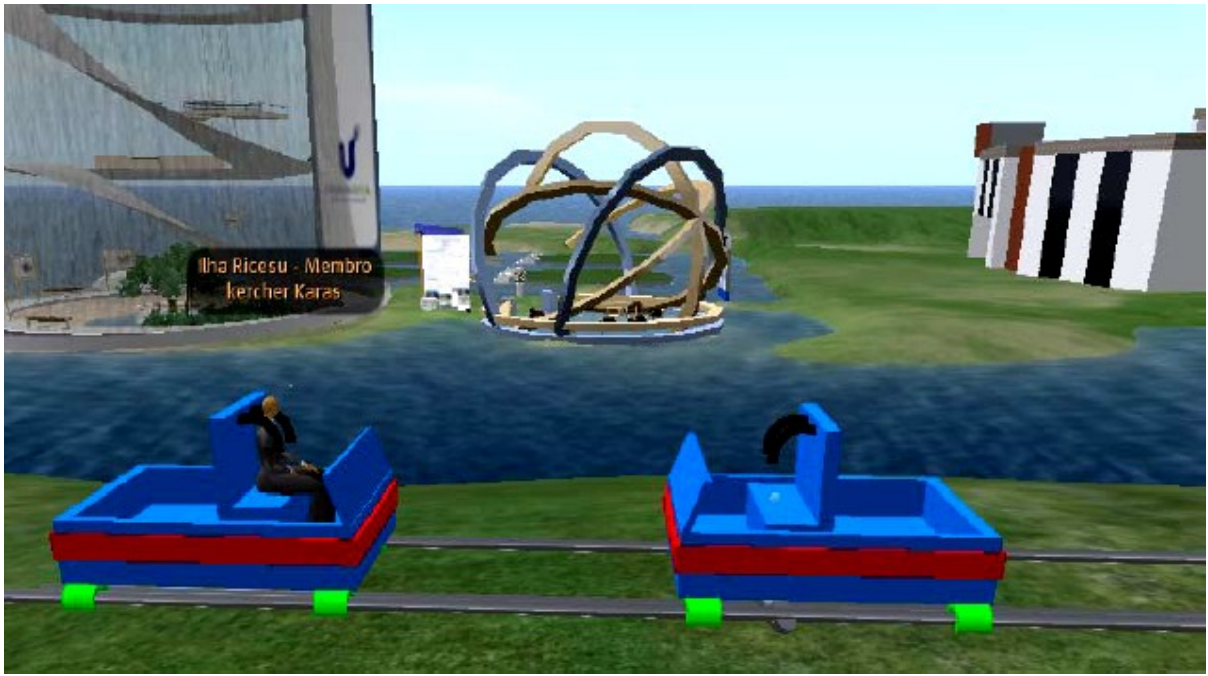

**Figure 2 - Bumper cars head-on collision simulator**
Source: (Greis & Reategui, 2010)

To investigate the feasibility of building physics microworlds in SL, we have built a 'cannon'[1] that can simulate both Newtonian Mechanics and its predecessor, Buridan's Impetus Theory as per user's choice. We have described in detail elsewhere (dos Santos, 2010a) our difficulties to counterbalance gravity and to achieve a rectilinear trajectory simulation for the cannonballs and our workaround for these difficulties.

---

[1] The operation of this cannon can be watched on video (dos Santos, 2010b, 2010c). It is freely accessible in our Second Life Physics Lab via the link http://www.fisica-interessante.com/virtual-second-life-physics-laboratory.html#Buridanian_Cannon

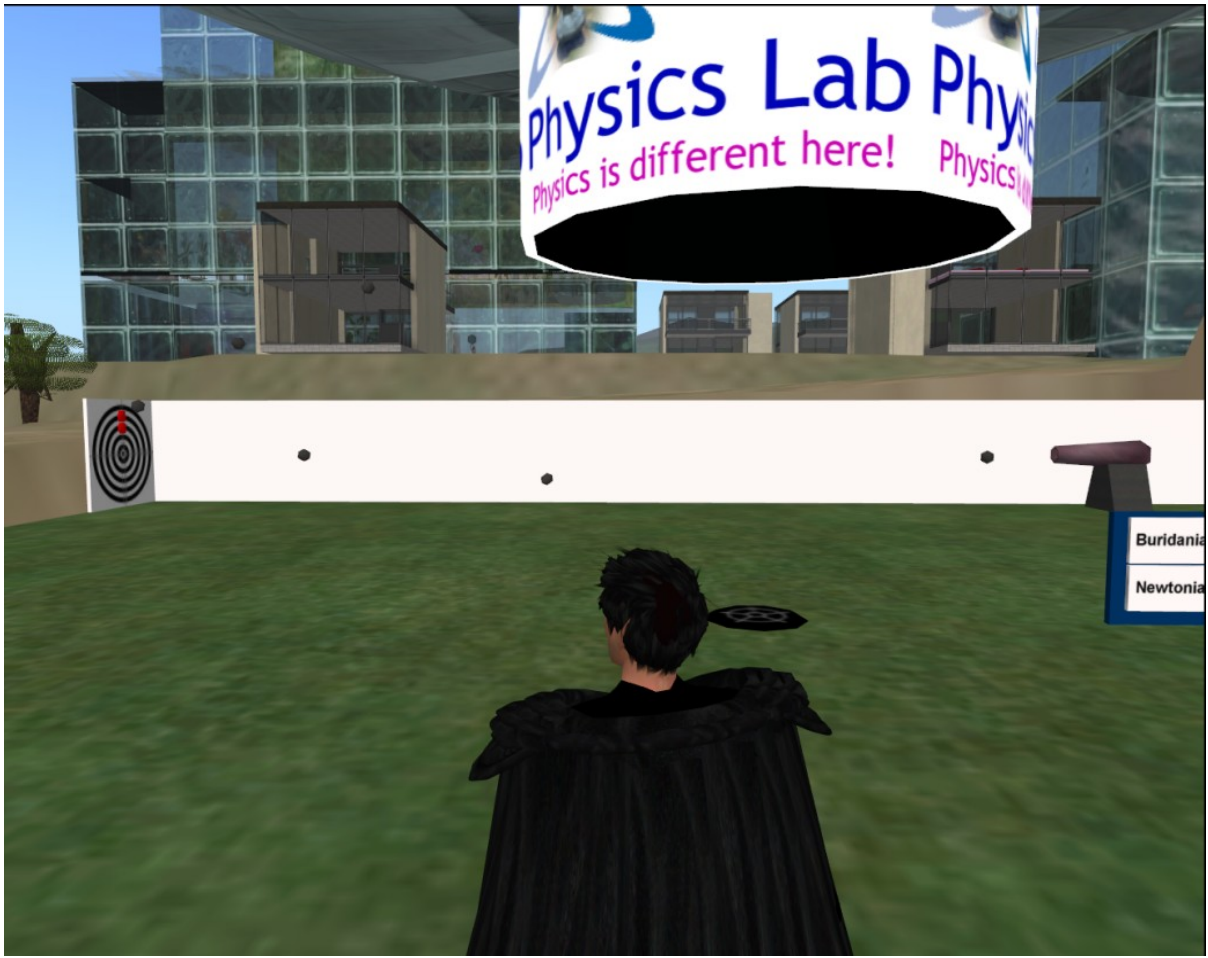

**Figure 3 - Projectiles falling down due to script delay to counterbalance gravity**
Source: photo taken by the author

Now, we proceed to evaluate the SL environment as a platform for physical microworlds and simulations.

**EVALUATION OF SL AS A PLATFORM FOR PHYSICAL SIMULATIONS**

Following Reis & Andrade Neto (2002), we will give more emphasis to features relevant to the teaching and learning process than to criteria more relevant in terms of computing like stability, aesthetics, compatibility, portability, and the like. We had, however, to generalize their criteria, as their work was focused its analysis on simulations of mechanical collisions.

**1 & 2. Representation of physical quantities in real time**
Physical quantities are not automatically represented by the SL client. Nevertheless, LSL provides functions to access directly some of them, e.g. *llGetPos* and *llGetVel* that return the position and velocity of the object in the region, respectively. Therefore, these quantities can be represented directly, in real time, during the entire duration of the phenomenon, as highlighted in Figure 1 and in Figure 4. There are no functions in LSL to access the value of some other relevant quantities, such as kinetic energy, momentum, and so forth that can be easily calculated by the script from those basic quantities, however.

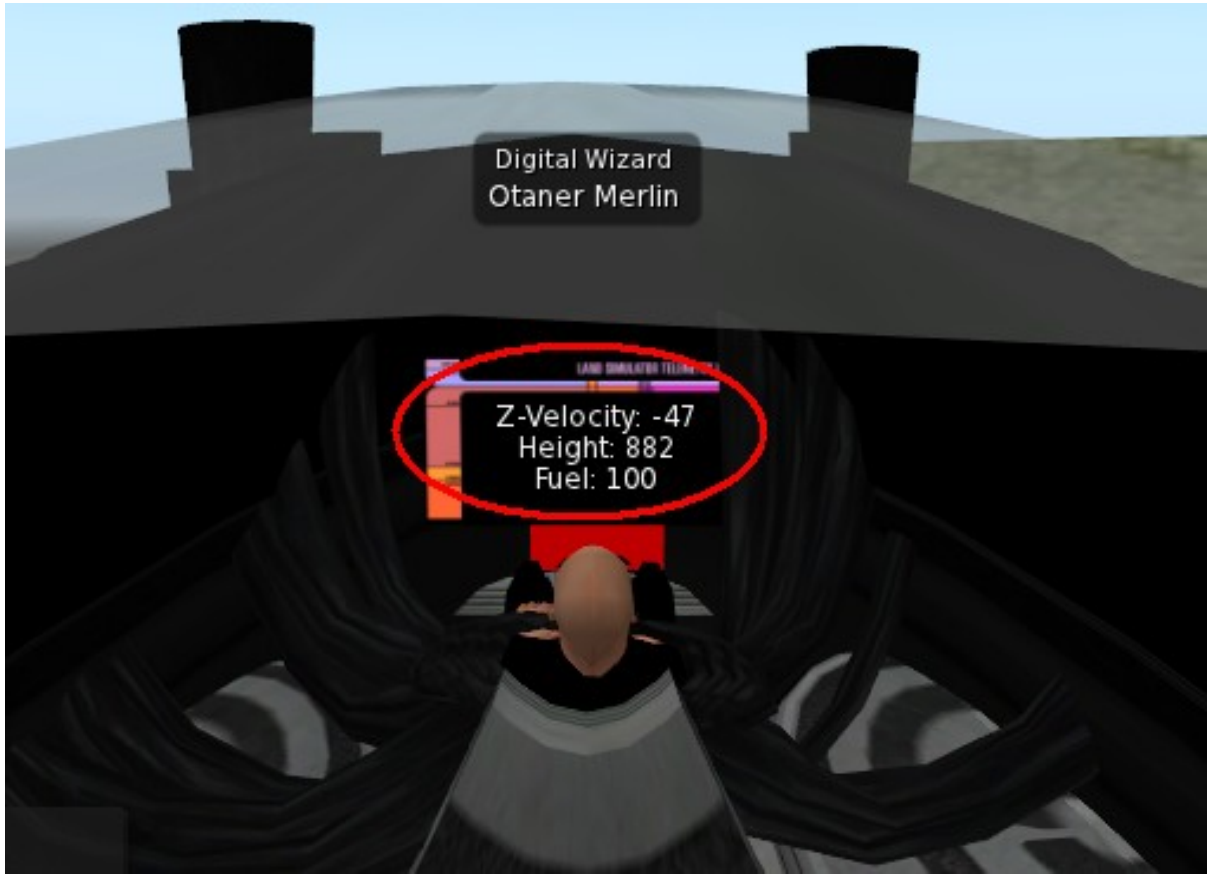

**Figure 4 – Information panel in our gamesim *Lander Simulator*[2].**
Source: photo taken by the author

**3. Explicit relationship between conservation of physical quantities and type of collision**
As we have seen above, the script either has access to the relevant quantities or can calculate them. Therefore, it is not a problem for the programmer to make the script to display some kind of warning, depending on the quantity is conserved or not. This could lead the student to pay attention to the way the experiment happened and make the desired association.

**4. Representation of physical quantities both in scalar and vector form**
LSL functions return the value of physical quantities in the correct scalar or vector form. We have already demonstrated elsewhere (dos Santos, 2012a) how the script can make use of the SL resources to represent them in correct form.

**5. Is it a simulation of an actual experiment?**
Reis & Andrade Neto (2002) argue that difficulties in student understanding, arising from differences between experiencing a phenomenon through an actual experiment or through a computer simulation, could be minimized if it represented an actual, easily accessible, phenomenon in student's everyday situations or laboratory. If we compare the collision simulators of Figure 1 and Figure 2, we have to agree that the former is more "real" in the sense that it duplicates a well-known laboratory equipment. However, the interactivity, the sense of 'presence', the participant observation from inside the experiment, of the second simulation, surely provides a richer, more intense, and meaningful experience to the student (Greis & Reategui, 2010). Likewise, the exploration of interactive life-size molecules (Lang & Bradley, 2009) and of fractals (Lee, 2007) do not represent 'real' experiments and much less accessible ones, but are innovative and stimulating experiences that definitely cannot be made in a classroom or in a laboratory. Therefore, we allow ourselves to disagree with Reis & Andrade Neto (2002) on this criterion, when referring to simulations in immersive 3D virtual worlds like SL.

---

[2] The operation of this *gamesim* can be watched on video (dos Santos, 2010d). It is freely accessible in our Second Life Physics Lab via the link http://www.fisica-interessante.com/virtual-second-life-physics-laboratory.html#Lander_Simulator

**6. Manipulation of initial parameters, e.g. velocities, masses, and positions**
This feature is very important because it assists students in understanding their relationship with the physical concepts and phenomena. It is feasible in SL, as the script can use the LSL *llRezObject* function to set the initial position and velocity of a physical object. To define the object's mass, it can use the *llSetScale* function, since in SL it depends on the size and not the 'material' of the object, as we have already demonstrated elsewhere (dos Santos, 2012b).

**7. Variety in visualization of physical quantities**
LSL and SL resources to build and manipulate textures and objects allow the script author to represent them in various forms, including bars, graphics, animations, icons.

**8. Ease of use of the program, stability, and portability**
The computational effort exerted by SL grid of computers leads to undesirable but understandable system instability and occasional falls. On the other hand, as almost all the processing occurs on servers, the client software is small and easily portable to a variety of operating systems. Unfortunately, however, the simulation itself cannot be exported as Java applets and be run remotely in a web browser as allowed by other platforms, e.g. *AnyLogic* (Rodrigues, 2008).

In conclusion, despite its rich resources, we cannot say that SL is an easy-to-use platform. Most users agree that a high learning curve exists for new users (Sanchez, 2009), which means that any proposal of using of SL for teaching has to set aside several hours, just to have the students become familiar with basic tasks, e.g. walking, pass through doorways, go up stairs, manipulate objects, and so forth.

**CONCLUSION**

After his study on students' issues in learning in SL, Sanchez (2009) concluded that "designers can create a user experience that will build on the strengths of the virtual world while overcoming the obstacles". We agree with Sanchez, and affirm that despite the restrictions and differences between the SL and a 'traditional' simulator, SL shows itself as a viable and flexible platform for microworlds and simulations. It surely requires some creativity to overcome the difficulties of implementation seen on the above examples that would be minimized if it had a friendlier interface for builders of simulations. On the other hand, the rich immersive 3D massively multi-user experience SL provides is pleasant, engages the user to explore the territory, and, therefore, offers a number of advantages over a 2D simulator. How much innovative, however, was Modellus (Teodoro et al., 1996) by providing the user with a very accessible interface, that eliminated the need to program graphic interfaces, at a time when most of the personal computer worked in DOS. Our vision is that SL is equally innovative as a 3D accessible platform.

Renato P. dos Santos
Associate Professor
PPGECIM (Doctoral Program in Science and Mathematics Education)
ULBRA - Lutheran University of Brazil
Av. Farroupilha 8001
92425-900 Canoas, RS
Brazil
Email: fisicaianteressante@gmail.com